\documentclass[11pt]{bmc_article_s50}

\usepackage{amssymb,amsmath}

\usepackage[]{natbib}
\usepackage{graphicx}
\usepackage{color}

\let\citet=\cite

\newcommand{\sub}[1]{_{\mbox{{\scriptsize #1}}}}
\newcommand \ga{\hspace{1ex} ^{>} \hspace{-2.5mm}_{\sim} \hspace{1ex}}
\newcommand \la{\hspace{1ex} ^{<} \hspace{-2.5mm}_{\sim} \hspace{1ex}}

\newcommand{\beq}{\begin{equation}}
\newcommand{\eeq}{\end{equation}}

\newcommand{\bea}{\begin{eqnarray}}
\newcommand{\eea}{\end{eqnarray}}

\newcommand{\rev}[1]{{#1}}

\usepackage{url}
\usepackage{setspace}
\usepackage[T1]{fontenc}
\usepackage{color}
\usepackage{ragged2e}
\justifying
\singlespace

\begin{document}

\title{Orbital evolution of planetesimals in gaseous disks}

\maketitle

\author[1*]{Hiroshi Kobayashi}\cor{}
\email{hkobayas@nagoya-u.jp}

\address[1]{Department of Physics, Nagoya University, Nagoya,
Aichi 464-8602, Japan}

\begin{abstract}
Planets are formed from collisional growth of small bodies in
a protoplanetary disk. Bodies much larger than approximately $1$\,m are mainly
controlled by the gravity of the host star and experience weak gas drag; their
orbits are mainly expressed by orbital elements: semimajor axes $a$,
eccentricities $e$, and inclinations $i$, which are modulated by gas
drag. In a previous study, $\dot a$, $\dot e$, and $\dot i$ were
analytically derived for $e \ll 1$ and $i \ll H/a$, where $H$ is the
scale height of the disk.  Their formulae are valid in the early stage
of planet formation.  However, once massive planets are formed, $e$ and
$i$ increase greatly.  Indeed, some small bodies in the solar system have
very large $e$ and $i$. Therefore, in this paper, I analytically derive formulae for
$\dot a$, $\dot e$, and $\dot i$ for $1-e^2 \ll 1$ and $i \ll H/a$ and
for $i \gg H/a$. The formulae combined from these limited
equations will represent the results of orbital integration unless $e
\geq 1$ or $i > \pi - H/a$. Since the derived formulae are applicable
for bodies not only in a protoplanetary disk but also in a circumplanetary
disk, I discuss the possibility of the capture of satellites in a
circumplanetary disk using the formulae.
\end{abstract}


\section*{Correspondence/Findings}

\subsection*{Introduction}

Planets are formed in a circumstellar disk composed of gas and
solid materials (solids are of the order of 1\% in mass).  The solid
material is initially sub-micron grains, which are controlled by an
aerodynamical frictional force that is much stronger than the gravity of
the central star \citep[][ hereafter AHN]{adachi}.  As solid bodies
grow, gas drag becomes relatively less important.  Once bodies get much
larger than 1\,m, they have Keplerian orbits around the central star
that are slightly altered by gas drag; then, their orbits are
characterized by orbital elements: semimajor axes $a$, eccentricities
$e$, and inclinations $i$.  These bodies grow via collisions, and the
collisional rates are given by relative velocities determined by $e$ and
$i$ \citep[e.g.,][]{inaba01}.  Damping due to gas drag and stirring by
the largest body in each annulus of the disk mainly control $e$ and $i$,
which evolve in the protoplanetary disk during planet formation. In addition, radial drift due to gas drag,
which is expressed by $\dot a$, reduces small bodies, which stalls the
growth of bodies \citep[e.g.,][]{kobayashi10,kobayashi11}.
Therefore, the time derivative of $a$, $e$, and $i$ ($\dot a$, $\dot e$, and $\dot i$) caused by gas drag are very
important for planet formation.

Protoplanets are formed out of collisions with kilometer-sized or larger
bodies called planetesimals.  While protoplanets grow, $e$ and $i$ of
planetesimals are determined by the Hill radius of the protoplanets, and
their $e$ and $i$ are smaller than 0.3 unless the protoplanets are greater
than ten Earth masses \rev{\citep[see equation 15 of][]{kobayashi10}}.
Therefore, AHN derived formulae of $\dot a$, $\dot e$, and $\dot i$ due to
gas drag for a body with low $e \la 0.3$ and $i \ll 0.1$.  However, $e$
and $i$ may possibly increase when more massive planets are
formed. Indeed, in the solar system, some comets, asteroids, and Kuiper
belt objects have very high $e$ and $i$ \citep[e.g.,][]{kobayashi05}.  In addition, if inclined and
eccentric orbits of irregular satellites around Jovian planets are
originated from captures due to interaction with circumplanetary disks
\citep[e.g.,][]{fujita}, these captured
bodies with high $e$ and $i$ evolve their orbits in the
disks. Therefore, analytic formulae for $\dot a$, $\dot e$, and
$\dot i$ for bodies with high $e$ and $i$ are helpful for the analysis
of small bodies in the late stage of planet formation.

In this paper, I first introduce a model for gaseous disks such as
protoplanetary and circumplanetary disks, and then, I revisit the derivation of
\citet{adachi} for the analytic formulae of $\dot a$, $\dot e$, and $\dot
i$.  Next, I derive $\dot a$, $\dot e$, and $\dot i$ for bodies
with high $e$ and/or high $i$.  By combining these limited solutions, I
construct approximate formulae for $\dot a$, $\dot e$, and $\dot i$,
which are applicable for all $e$ and $i$ unless $e \geq 1$ or $i > \pi -
H/a$. Lastly, I discuss the orbital evolution
of satellites captured by circumplanetary disks
using the derived analytic formulae for $\dot a$, $\dot e$, and $\dot
i$.

\subsection*{Nebula disk model and gas drag law}\label{sc:model}

In order to evaluate the drag force due to nebula gas, the
disk model is set as follows.  A gaseous disk rotates around a
central object with mass $M_*$, which is axisymmetric and in a steady
state.
In a cylindrical coordinate system ($r,\theta,z$), the gas density $\rho$ is
defined from the force equilibrium in the $z$ direction in a vertical
isothermal disk as
\begin{equation}
\rho = \frac{\sigma}{\sqrt{\pi} H(r)} \exp \left( - \frac{z^2}{H(r)^2} \right),
\label{eq:rho}
\end{equation}
where $\sigma(r) ( = \int_{-\infty}^{\infty} \rho dz)$ is the surface
density of the nebula disk, $H(r) = \sqrt{2} c/\Omega_{\rm K}$ is the
disk scale height, $\Omega_{\rm K} = (G M_* / r^3)^{1/2}$ is the Keplerian
angular velocity, and $G$ is the gravitational constant.  For
simplicity, the $r$-dependences of $\sigma$ and $c$ are assumed as
$\sigma \propto r^{-\alpha}$, $c \propto r^{-\beta}$, respectively. This
relations give $\rho \propto r^{-\alpha^{\prime}}$, where
$\alpha^{\prime} = \alpha -\beta + 3/2$.  In the minimum-mass solar
nebula model \citep{hayashi85}, for example, $\alpha = 3/2$ and $\beta =
1/4$.  The angular gas velocity $\Omega_{\rm gas}$ is obtained from the
force equilibrium in the $r$ direction as \citep{tanaka}
\begin{equation}
\Omega_{\rm gas} = \Omega_{\rm K}
    \left[ 1 - \frac{1}{4} \left(\alpha + \beta +\frac{3}{2}\right)
    \frac{H(r)^2}{r^2} - \frac{\beta}{2} \frac{z^2}{r^2}
    \right].
\label{eq:anglar}
\end{equation}
In Equation (\ref{eq:anglar}), the terms of ${\cal O}(z^4/r^4)$ and
higher are ignored.  This treatment is valid even for investigation of the gas
drag effect on highly inclined orbits because the gas drag (and the
nebula gas) is negligible at a high altitude ($z \gg H$).

At the midplane of the disk, the relative velocity difference between
the gas motion and the Keplerian rotation is given by
\begin{equation}
 \eta(r) = \left. \frac{\Omega_{\rm K} -\Omega_{\rm gas}}{\Omega_{\rm K}}
    \right|_{\rm z = 0} = \frac{1}{4} \left(\alpha + \beta +
    \frac{3}{2}\right) \frac{H(r)^2}{r^2}.
\end{equation}

For a body with mass $m$ and radius $d$, gas drag force per unit mass
can be written as AHN
\begin{equation}
\textbf{F}_{\rm d} = C_{\rm D} \pi d^2 \rho u \textbf{u} / 2 m = A \rho u \textbf{u},
\label{eq:force}
\end{equation}
where $C_{\rm D}$ is the dimensionless gas drag coefficient, $\textbf{u}$
is the relative velocity vector between the body and the gas, $u =
\mid \textbf{u} \mid$, and $A = C_{\rm D} \pi d^2/2m$.
Although $C\sub{D}$ depends on Mach number $M$ and
Reynolds number $Re$, $C_{\rm D}$ is a constant
for high $Re$ ($d \ga 1$\,km in the minimum-mass solar nebula)
or for $M \gg 1$ ($e$ or $i$ is much larger than $H/a$) (AHN).

\subsection*{General expressions for the change in $\boldsymbol{a}$, $\boldsymbol{e}$, and $\boldsymbol{i}$}\label{sc:derive}

In this paper, I investigate the time variations of semimajor axis $a$,
eccentricity $e$, and inclination $i$ of a body due to gas drag for constant
$C_{\rm D}$ (and then constant $A$).  The time derivatives of $a$, $e$,
and $i$ are given by AHN as
\begin{eqnarray}
 \frac{d a}{d t} &=&
- A \rho u \frac{2a}{1-e^2} \left[ 1 + 2e \cos f + e^2
- (1+ e \cos f)^{3/2} \kappa \cos i \right],
\label{eq:gauss_a}
\\
\frac{de}{dt} &=& - A \rho u
 \left[2 \cos f + 2 e
  - \frac{2 \cos f + e + e \cos^2 f}{(1+e \cos f)^{1/2}}\kappa \cos i
 \right],
\\
\frac{di}{dt} &=&
 - A \rho u \frac{\cos^2 (f + \omega)}{(1+e \cos f)^{1/2}}\kappa
 \sin i,
\label{eq:gauss_i}
\end{eqnarray}
where
$f$ and $\omega$ are the true anomaly and  the argument of pericenter,
respectively,
$\kappa = \Omega_{\rm gas}/\Omega_{\rm K} w^{3/2}$,
$w = [1 -\sin^2 (f+\omega) \sin^2 i]^{1/2}$,
\begin{eqnarray}
\rho &=&
\displaystyle
\rho_0 \left( \frac{(1-e^2) w}{1+e \cos f}\right)^{-\alpha+\beta - 3/2}
\exp \left(
- \frac{a^2 (1-e^2)^2 w^2 \sin^2 (\omega + f) \sin^2 i}{H(r)^2 (1+ e \cos f)^2}
\right),
\label{eq:rho_new}
\\
u
&=& \frac{v_{\rm K}(a)}{\sqrt{1-e^2}}[
    1+2 e \cos f + e^2
- 2 ( 1 + e \cos f)^{3/2} \kappa \cos i
    + ( 1 + e \cos f) \kappa^2 w^2 ]^{1/2},
\label{eq:u}
\end{eqnarray}
\rev{$\rho_0$ is the midplane density at $r = a$, and $v_{\rm k} = (G
M_* / a)^{1/2}$ is the Keplerian velocity.
}
If the variation timescales of $a$, $e$, and $i$ are much longer than the
orbital time, the evolution of $a$, $e$, and $i$ follows the averaged
rate.  The orbital averaging is taken as
\begin{equation}
\left< \frac{da}{dt} \right> =
    \frac{1}{T_{\rm K}} \int^{T\sub{K}}_{0} \frac{da}{dt} dt
    = \frac{1}{2\pi} \int^{2 \pi}_{0} \frac{da}{dt}
            \frac{(1-e^2)^{3/2}}{(1+e\cos f)^2} df,\label{eq:average}
\end{equation}
\rev{where $T_{\rm K} = 2 \pi ( a^3 / G M_*)^{1/2} $ is the Keplerian period.}
The same averaging is taken for $e$ and $i$.

\subsection*{Case of low $\boldsymbol{e, i,}$ and $\boldsymbol{\eta}$}

For $e, i \ll 1$, \citet*{adachi}  derived the averaged changes in
$a$, $e$, and $i$ for three cases, (i) $\eta \gg e, i$, (ii) $i \gg
e,\eta$, and (iii) $e \gg i,\eta$, and summed up the leading terms for
these cases.
\rev{
This method was used to treat $u$ in
Equations~(\ref{eq:gauss_a}) to (\ref{eq:gauss_i}) analytically: The assumptions
simplify as
$u \approx \eta + (e/2) \cos f$  in case (i),
$u \approx i \mid \cos (f+\omega) \mid$ in case (ii), and
$u \approx e \sqrt{1- (3/4) \cos^2 f} + (\eta/2) \cos f  \sqrt{1- (3/4)
\cos^2 f}$ in case (iii). Other terms are also simplified, such as
$\rho = \rho_0 ( 1 + \alpha^{\prime} e \cos f)$. Then, the terms are
easily averaged over the orbital period by Equation~(\ref{eq:average}).
}

The derived formulae are in good agreement with the results of orbital
integrations for $e \ll 1$ and $i \ll H(a)^2/a^2$.
While \citet*{adachi} provided the term of $i^3$ in $\dot a$,
they did not take into account the vertical dependence of $\rho$, which
includes other $i^3$ terms. Since the sum of these $i^3$ terms is
negligible, I thus exclude the $i^3$ term derived by AHN.
\citet{inaba01} found that the mean root squares of these limited
solutions are in better agreement with the results of orbital evolution
than the simple summation by \citet*{adachi}.
The averaged variation rates of $a$, $e$, and $i$ are therefore given by
\begin{eqnarray}
- \frac{\tau_0} {a}\left\langle \frac{da}{dt} \right\rangle_1  &=&
2 \left\{
\eta^4 + \left(\frac{2i\eta}{\pi}\right)^2
+ \left[
   \frac{2(2E+K)}{3\pi} e \eta + \left(\frac{2E+K}{9\pi} \alpha^{\prime}
                  + \frac{68E-11K}{54\pi}\right) e^3
\right]^2
\right\}^{1/2},
\label{eq:mod_adachi_a}
\\
-\frac{\tau_0}{e}
\left\langle \frac{de}{dt} \right\rangle_1 &=&
\left[
\left(
\frac{3}{2}\eta\right)^2
+
\left(
\frac{2}{\pi}i
\right)^2
+
\left(
\frac{2E}{\pi}e
\right)^2
\right]^{1/2},
\\
-
\frac{\tau_0}{i}
\left\langle
\frac{di}{dt}
\right\rangle_1
&=& \frac{1}{2}
\left\{
\eta^2
+
\left(
\frac{8}{3\pi}i
\right)^2
+
\left[
\frac{2E}{\pi}e \left( 1+\frac{2K-5E}{9E} \cos 2\omega \right)
\right]^2
\right\}^{1/2},
\label{eq:mod_adachi_i}
\end{eqnarray}
where $K=2.157$ and $E=1.211$ are the first and second complete elliptic
integrals of argument $\sqrt{3/4}$, respectively, and $\tau_0 = (A \rho_0 v_{\rm
K})^{-1}$ is the stopping time due to gas drag for $u = v_{\rm K}$.
Note that I corrected an error in the factor of the $\eta^2$ term for $\dot e$ in
\citet{adachi}, which was pointed out by \citet{kary}.

\begin{figure}[!h]
\vskip-5pt
\hrulefill
\begin{center}
\includegraphics[width=10cm]{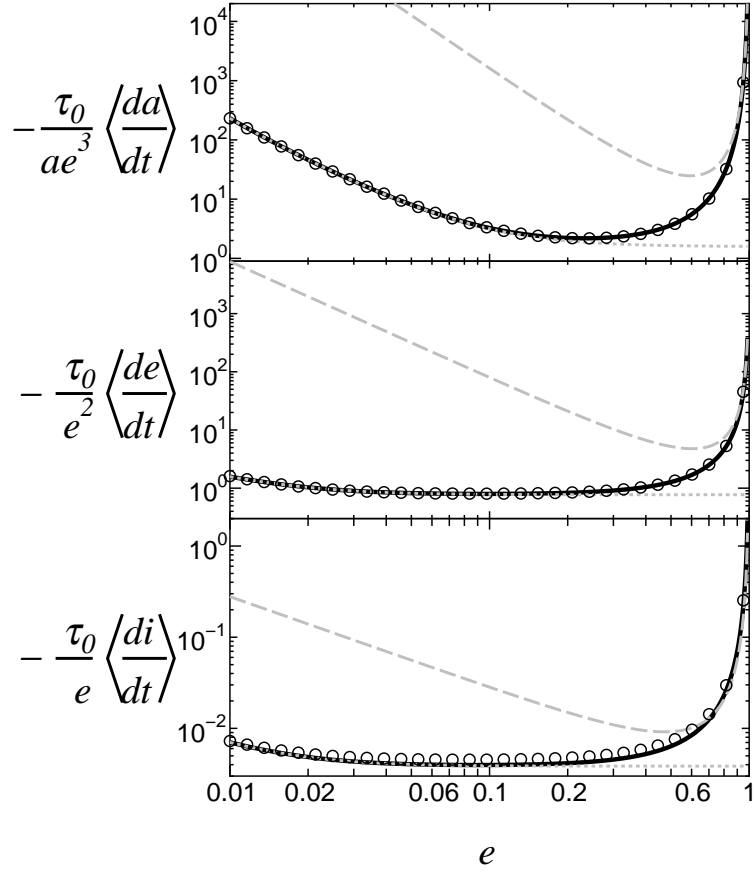}
\end{center}
\caption{ 
The variation rates of $a$, $e$, and $i$ as a function of $e$
for 
$i = 0.01$
and 
$\omega$ $=$ $\pi/2$ 
in the disk with $H(a)/a = 0.1$,
$\alpha = 1.5$, and $\beta = 0.25$.  
Analytic formulae for low $e$ 
(gray dotted curves), given by Equations~(\ref{eq:mod_adachi_a}) to
(\ref{eq:mod_adachi_i}), and ones for high $e$ (gray dashed curves),
given by Equations~(\ref{eq:da_he}) to (\ref{eq:di_he}), are in good
agreement with the results of orbital integration (open circles) for low
$e$ or high $e$, respectively. The combined formulae (solid curves),
given by Equations~(\ref{eq:da_high}) to (\ref{eq:di_mid}), are valid
for the whole region.}  \label{fig1}
\vskip-5pt
\hrulefill
\end{figure}

\begin{figure}[!h]
\vskip-5pt
\hrulefill
\begin{center}
\includegraphics[width=10cm]{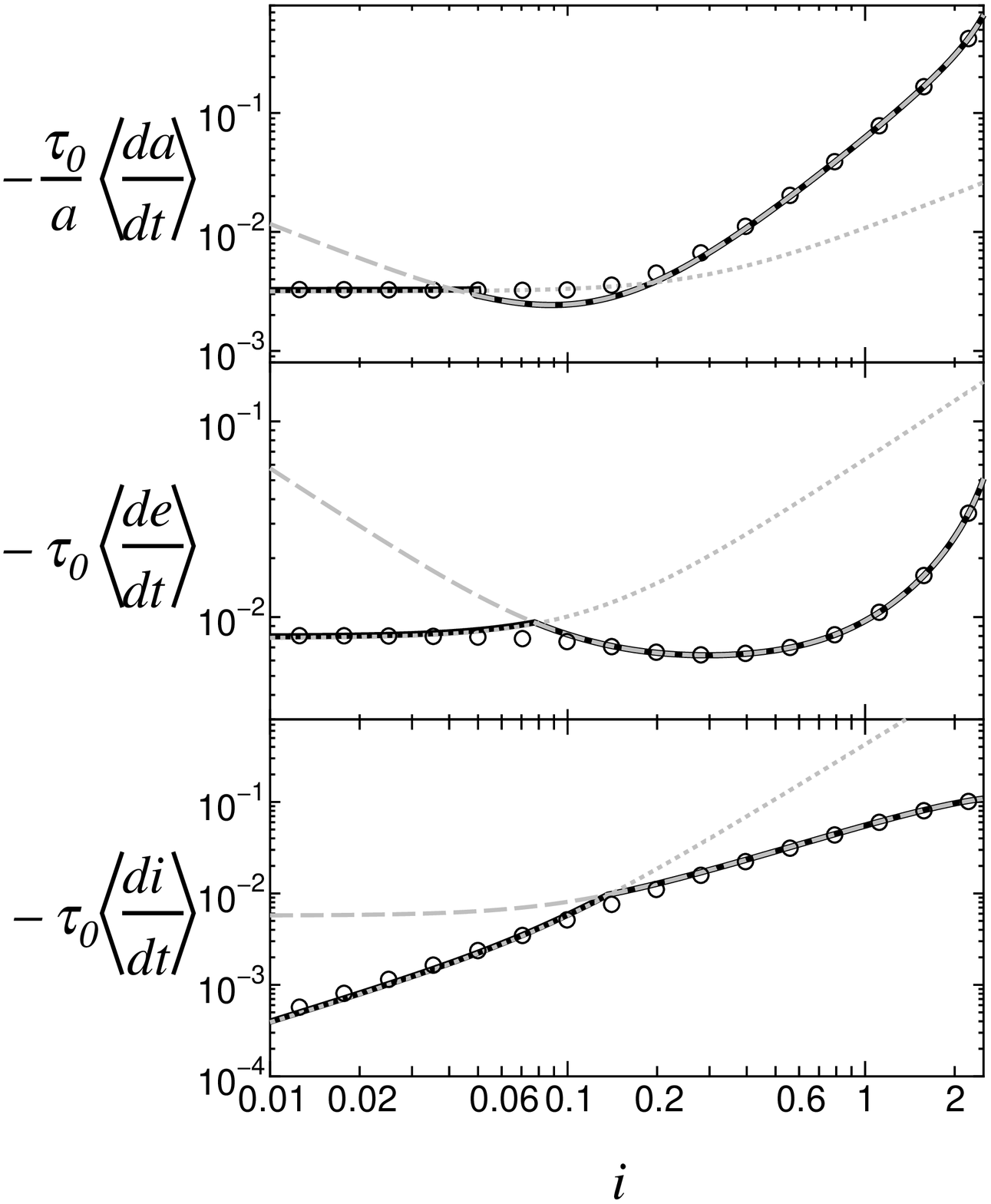}
\end{center}
\caption{The change rates in $a$, $e$, and $i$ as a function of $i$
for $e = 0.1$, and $\omega = \pi/2$ in the same disk as
 Figure~\ref{fig1}.
Analytic formulae for low $i$ (gray dotted curves), given by
 Equations~(\ref{eq:mod_adachi_a}) to (\ref{eq:mod_adachi_i}),
and those for high $i$ (gray dashed curves), given by
 Equations~(\ref{eq:da_high}) to (\ref{eq:di_high}), are in good agreement with
the results of orbital integration (open circles) for $i$ $\ll H/a$ and
 $i$ $\gg H/a$, respectively. The combined formulae (solid curves), given by
 Equations~(\ref{eq:da_high}) to (\ref{eq:di_mid}), represent within a factor of
 $1.5$.}
\label{fig2}
\vskip-5pt
\hrulefill
\end{figure}

For $i=0.01$, Equations~(\ref{eq:mod_adachi_a}) to (\ref{eq:mod_adachi_i}) are
compared with the results of orbital integrations in
Figure~\ref{fig1}. These formulae are valid unless $e > 0.2$. Moreover, the $i$ dependence in these formulae are valid for $i <
H(a)/a$ (see Figure~\ref{fig2}).

\subsection*{Case of high eccentricity and low inclination}

Here, let us consider the case where $e$ is almost equal to unity and
$i$ is much smaller than $H(a)/a$.  Expanding
Equations (\ref{eq:gauss_a}) to (\ref{eq:gauss_i}) with respect to $(1-e^2)$
under the assumption of $i \ll H(a)/a$, keeping only the lowest-order
terms of $(1-e^2)$, and applying the orbital averaging such as
Equation~(\ref{eq:average}) to these equations,
\begin{eqnarray}
 \left< \frac{da}{dt} \right> &=& - \frac{2 a}{\tau_0}
  (1-e^2)^{-\alpha+\beta-3/2} \Psi,
\label{eq:da_he}
\\
\left< \frac{de}{dt} \right> &=& - \frac{1}{\tau_0}
 (1-e^2)^{-\alpha+\beta-1/2} \Psi,
\label{eq:de_he}
\\
\left< \frac{di}{dt} \right> &=& - \frac{i}{2 \tau_0}
 (1-e^2)^{-\alpha+\beta-1/2} \Phi_1
\left(
1 + \frac{\Phi_2}{\Phi_1} \cos 2 \omega
\right),
\label{eq:di_he}
\end{eqnarray}
where
\begin{eqnarray}
  \Psi
&=& \frac{1}{2\pi}\int^{2 \pi}_{0} (1 + \cos f )^{\alpha-\beta+1}(2-\sqrt{1+ \cos f})
    \sqrt{3-2\sqrt{1+ \cos f}} df,
\\
 \Phi_1 &=& \frac{1}{2\pi}\int^{2 \pi}_{0}
(1 + \cos f )^{\alpha-\beta-1/2}\sqrt{3-2\sqrt{1+ \cos f}}df,
\\
 \Phi_2 &=& \frac{1}{2\pi} \int^{2 \pi}_{0}
(1 + \cos f )^{\alpha-\beta-1/2}\sqrt{3-2\sqrt{1+ \cos f}} \cos 2 fdf.
\end{eqnarray}
\rev{
The dependences of $\dot a$ and $\dot e$ on $f$ are seen in the integral
$\Psi$, while a term proportional to  $\sin 2 f \sin 2
\omega$ in $\dot i$ vanishes by the orbital averaging because of
an odd function of $f$.
}
The integrals of $\Psi$, $\Phi_1$, and $\Phi_2$
are functions of $\alpha - \beta$.  In the minimum-mass solar nebular
model, $\alpha-\beta$ is 5/4, and then, $\Psi = 0.79$, $\Phi_1 = 0.71$,
and $\Phi_2 = -0.16$.

The $e$ dependences in these formulae are applicable for $e>0.9$ as shown
in Figure~\ref{fig1}. Although the effective range of these formulae
is limited, the $e$ dependences improve the high $e$ parts in
Equations~(\ref{eq:mod_adachi_a}) to (\ref{eq:mod_adachi_i}) as shown below.

\subsection*{Case of high inclination}

Next, let us consider highly inclined orbits where $a i / H(a) $ is much
larger than unity.  Bodies with such a high inclination penetrate the
nebula disk twice around the ascending and descending nodes through an
orbital period. Gas drag is effective only around the nodes.
Since the body experiences significant
gas drag around the ascending node ($\mid f+\omega \mid \ll 1$),
the leading terms of $\mid f+\omega
\mid$ for Equations (\ref{eq:gauss_a}) to (\ref{eq:gauss_i}) are
\begin{eqnarray}
\frac{da}{dt} &=& -\frac{2 a}{\tau_0 (1-e^2)^{3/2} \tilde r^{-\beta + 5/2}} I(\omega) \exp \left(
- \frac{a^2 \tilde r^2 (f+\omega)^2 \sin^2 i}{H(r)^2 }
\right),\label{eq:high_i_a}
\\
\frac{de}{dt} &=& -\frac{1}{\tau_0 \sqrt{1-e^2} \tilde r^{-\beta + 5/2}} J(\omega)
\exp \left(
- \frac{a^2 \tilde r^2 (f + \omega)^2 \sin^2 i}{H(r)^2 }
\right),
\\
\frac{di}{dt} &=& -\frac{\sin i}{\tau_0 (1-e^2)^{3/2} \tilde r^{-\beta + 5/2}} K(\omega)
\exp \left(
- \frac{a^2 \tilde r^2 (f+\omega)^2 \sin^2 i}{H(r)^2}
\right),\label{eq:high_i_i}
\end{eqnarray}
where
\begin{eqnarray}
I(\omega) &=& \tilde{r}^{-\alpha + 1}
    \tilde{u} [ 1 + 2 e \cos \omega + e^2
        - ( 1 + e \cos \omega)^{3/2} \cos i],
\\
J(\omega) &=& \tilde{r}^{-\alpha + 1}
        \tilde{u} \left[ 2( e + \cos \omega)
- \left( \cos \omega + \frac{\cos \omega +e}{1 + e\cos \omega} \right)
\cos i \sqrt{1+e \cos \omega} \right],
\\
K(\omega) &=& \tilde{r}^{-\alpha + 2}
    \tilde{u} \sqrt{1 + e \cos \omega},
\end{eqnarray}
and
\begin{eqnarray}
\tilde{r} &=& \frac{1-e^2}{1+ e \cos \omega},
\\
\tilde{u} &=& \sqrt{2 + 3 e \cos \omega + e^2 - 2 ( 1 + e \cos \omega)^{3/2} \cos i}.
\end{eqnarray}
For this derivation, $\Omega_{\rm gas} = \Omega_{\rm K}$, since the
relative velocity is mainly determined by inclination.

\rev{
In order to apply
averaging over half an orbit around the ascending node, $\dot a$, $\dot
e$, and $\dot i$ are integrated from $f = - \omega - \pi/2 $ to $f =
-\omega + \pi/2$.
Since
$\dot a$, $\dot e$, and $\dot i$ are Gaussian functions as shown in
Equations~(\ref{eq:high_i_a}) to (\ref{eq:high_i_i}), they are negligible for
large $(f+\omega)^2$ and the integral is thus approximated to be that over
interval [$-\infty,\infty$] as follows:
}

\begin{eqnarray}
\int^{-\omega+\pi/2}_{-\omega-\pi/2}
\exp \left(
- \frac{a^2 \tilde r^2 (f+\omega)^2 \sin^2 i}{H(r)^2}
\right) df
& \simeq &
\int^{\infty}_{-\infty}
\exp \left(
- \frac{a^2 \tilde r^2 (f+\omega)^2 \sin^2 i}{H(a \tilde r)^2}
\right) d(f+\omega),
\\
&=& \frac{H \tilde r^{-\beta + 1/2} \sqrt{\pi}}{a \sin i},
\end{eqnarray}

where $H = H(a)$.  Using this,
Equations (\ref{eq:high_i_a}) to (\ref{eq:high_i_i}) are integrated around the
ascending node, which results in the averaged variation rates of $a$, $e$,
and $i$ in half an orbit.

The variation rates due to the penetration near the descending node ($f
\approx - \omega - \pi$) are obtained in the same way as above.  Summing up the
changes at two penetrations, the averaged changes are given by
\begin{eqnarray}
\left< \frac{da}{dt} \right>_{\rm high} &=& -  \frac{a}{\tau_0} \frac{H}{\sqrt{\pi} a (1-e^2)^2 \sin i}
    [I(\omega) + I(\omega + \pi)],\label{eq:da_high}
\\
\left< \frac{de}{dt} \right>_{\rm high} &=& - \frac{1}{\tau_0}
\frac{H}{2 \sqrt{\pi} a (1-e^2)  \sin i}
    [J(\omega) + J(\omega + \pi)],
\\
\left< \frac{di}{dt} \right>_{\rm high} &=& - \frac{1}{\tau_0}
\frac{H}{2\sqrt{\pi}a (1-e^2)^2}
    [K(\omega) + K(\omega + \pi)].\label{eq:di_high}
\end{eqnarray}

\begin{figure}[!h]
\hrulefill
\vskip-5pt
\begin{center}
 \includegraphics[width=10cm]{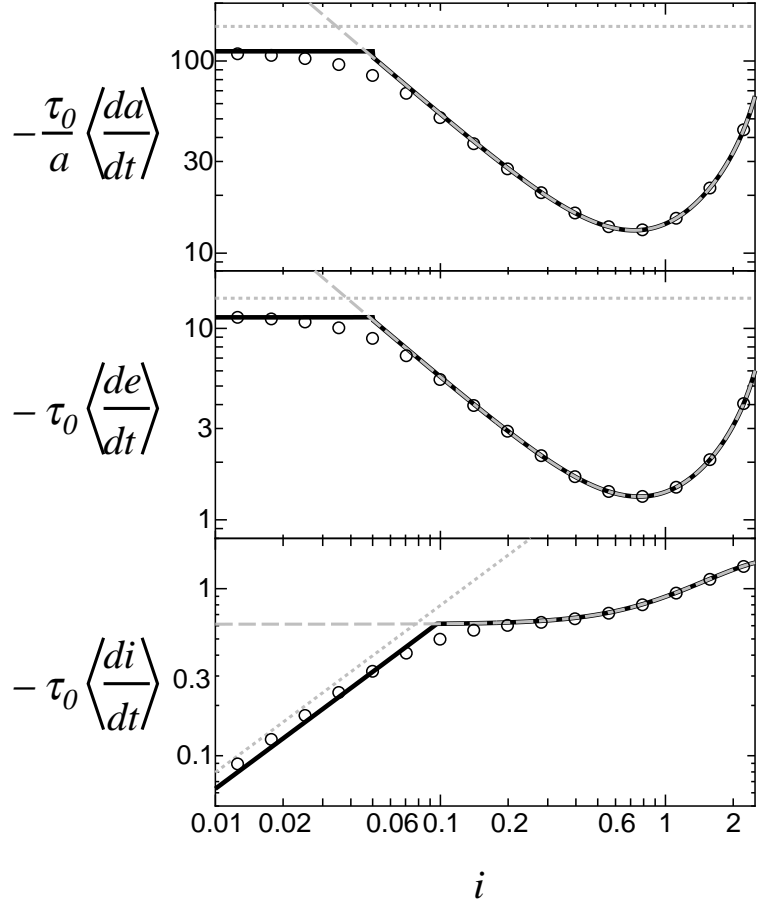}
\end{center}
\caption{Same as Figure~\ref{fig2}, but for $e=0.9$ and dotted lines given by
 Equations~(\ref{eq:da_he}) to (\ref{eq:di_he}).}
\label{fig3}
\vskip-5pt
\hrulefill
\end{figure}

The validity of Equations~(\ref{eq:da_high}) to (\ref{eq:di_high}) is shown in
Figures~\ref{fig2} and \ref{fig3}. These formulae are
applicable for $i > 2H/a$.

\subsection*{Combined equations}

The variation rates of $a$, $e$, and $i$ in two limited cases for $i \ll
H/a$ are derived above. The formulae for low $e$ do not well reproduce
the variation rate in $e \sim 1$, while high-$e$ formulae overestimate
the values for low $e$.  Combination of low-eccentricity formulae of
Equations~(\ref{eq:mod_adachi_a}) to (\ref{eq:mod_adachi_i}) with the $1-e^2$
dependence derived in Equations~(\ref{eq:da_he}) to (\ref{eq:di_he}) gives
\begin{eqnarray}
 \left\langle \frac{da}{dt} \right\rangle_{\rm low} &=&
  \left\langle \frac{da}{dt} \right\rangle_1 (1-e^2)^{-\alpha + \beta
  -3/2},\label{eq:da_low}
  \\
\left\langle \frac{de}{dt} \right\rangle_{\rm low} &=&
 \left\langle \frac{de}{dt} \right\rangle_1
  (1-e^2)^{-\alpha + \beta -1/2},
  \\
\left\langle \frac{di}{dt} \right\rangle_{\rm low} &=&
 \left\langle \frac{di}{dt} \right\rangle_1
 (1-e^2)^{-\alpha + \beta -1/2}.
\label{eq:di_low}
\end{eqnarray}
These formulae are given in a very simple way, but they are
in good agreement with the results of orbital integration if
$i < H/2a$ (see Figures~\ref{fig1} to \ref{fig3}).

If $H/2a <i < H/a$, the variation rates of $a$, $e$, and $i$ are
obtained from combination of the low-$i$ formulae of
Equations~(\ref{eq:da_low}) to (\ref{eq:di_low}) and the high-$i$ formulae of
Equations~(\ref{eq:da_high}) to (\ref{eq:di_high}).
\begin{eqnarray}
 \left\langle \frac{da}{dt} \right\rangle_{\rm mid} &=&
  {\rm MIN}\left(\left\langle\frac{da}{dt} \right\rangle_{\rm low}
       ,\left\langle\frac{da}{dt} \right\rangle_{\rm high}
       \right),\label{eq:da_mid}
\\
 \left\langle \frac{de}{dt} \right\rangle_{\rm mid} &=&
  {\rm MIN}\left(\left\langle\frac{de}{dt} \right\rangle_{\rm low}
       ,\left\langle\frac{de}{dt} \right\rangle_{\rm high}
       \right),
\\
 \left\langle \frac{di}{dt} \right\rangle_{\rm mid} &=&
  {\rm MIN}\left(\left\langle\frac{di}{dt} \right\rangle_{\rm low}
       ,\left\langle\frac{di}{dt} \right\rangle_{\rm high}
       \right),\label{eq:di_mid}
\end{eqnarray}
where ${\rm MIN}(D,E)$ is the smaller of $D$ and $E$.

In conclusion, the variation rates for $a$, $e$, and $i$ are
approximately given by
\begin{itemize}
 \item Equations~(\ref{eq:da_low}) to (\ref{eq:di_low}) for $i \leq H/2a$,
 \item Equations~(\ref{eq:da_mid}) to (\ref{eq:di_mid}) for the intermediate
       inclination ($H/2a < i \leq 2 H/a$),
\item Equations~(\ref{eq:da_high}) to (\ref{eq:di_high}) for $i > 2H/a$.
\end{itemize}
In the intermediate $i$, the formulae tend to deviate from the right
values but the accuracies are within a factor of 1.5 (see
Figures~\ref{fig1} to \ref{fig3}).
It should be noted that these formulae are not applicable to
the case of $i > \pi - H/a$ where a body experiences gas drag with relative
velocity $\sim v_{\rm K}$ not only around the nodes but also for a whole orbit.

\subsection*{Application to captured satellites}

Jovian planets have many satellites, which may be formed in circumplanetary
disks. Satellites close to planets mainly have circular and coplanar
orbits and may be formed in the disks. However, distant satellites tend
to have inclined orbits. Here, I discuss the possibility of the capture of
satellites in the disks because the formulae for $\dot a$, $\dot e$, and
$\dot i$ that I derive in this paper are applicable to bodies with
high $e$ and $i$.

Orbital evolution of bodies with high $e$ is predicted from these
analytic formulae. When a body is captured by gas drag in a
circumplanetary disk, $e$ of the captured body is approximately $1$. For $e >
0.9$, $|\dot e|/e$ and $|\dot a|/a$ are very large. Variation rate of
the pericenter distance $q$ is much smaller than those of $a$ and
$e$. Indeed, $\dot q = (1-e) \dot a - a \dot e$ is estimated to be zero
in Equations~(\ref{eq:da_he}) and (\ref{eq:de_he}).  The result is caused
by the neglect of the higher terms of $(1-e^2)$, and these higher
$(1-e^2)$ terms give $\dot q/q$ a positive value but $\dot q/q$ is much
smaller than $|\dot a|/a$ and $|\dot e|/e$.  Therefore, the orbital
evolution occurs along with almost constant $q$.  With decreasing $e$, the
orbital evolution changes.  Since $|\dot a|/a$ becomes smaller than
$|\dot e|/e$ for $e < 0.5$ to 0.6, $e$ decreases with almost constant $a$.
Once $e \ll \eta$, $\dot a$ becomes dominant for the orbital evolution; the
body drifts to the host planet in the timescale of $\tau_0 / 2 \eta^2$.

\begin{figure}[!h]
\hrulefill
\vskip-5pt
\begin{center}
   \includegraphics[width=10cm]{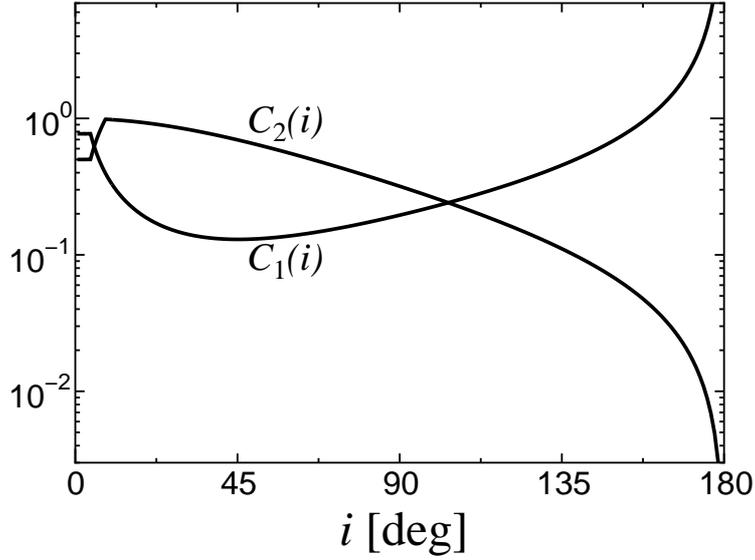}
\end{center}
\caption{  $C_1(i) \equiv (\Delta e)_{e =1} \, \tau_0(q)/T_{\rm K}(q)$
 and $C_2(i)  \equiv (\langle \dot i \rangle /i \langle \dot e \rangle
  )_{e =1}$ derived from the combined
  equations (Equations~\ref{eq:da_he} to \ref{eq:di_he}).}
\label{fig4}
\vskip-5pt
\hrulefill
\end{figure}

The bodies that will be satellites are temporally captured by a planet at first
\citep{suetsugu11,suetsugu13}, and the apocenter distances of the bodies
decrease to less than the Hill radius of the host planet during the
temporal capture of bodies \citep[e.g.,][]{fujita}.  The change of
orbital eccentricity in an orbit around the host planet is given by $\Delta
e \approx \langle \dot e \rangle T_{\rm K}$.
The body is fully captured by gas drag if $f_{\rm cap}
\Delta e \sim 1$ during the temporal capture, where $f_{\rm cap}$ is the
number of close encounters with the planet during the temporal capture.
Using the combined formulae (Equations~\ref{eq:da_high} to \ref{eq:di_mid}) at
$e=1$, $\Delta e$ is given by $C_1 (i) T_{\rm K}(q) /\tau_0(q)$, where
$T_{\rm K}(q)$ and $\tau_0(q)$ are $T_{\rm K}$ and $\tau_0$ at the
pericenter distance $q$, respectively. Therefore, the necessary
condition for capture is given by
\begin{equation}
 \rho \ga 4 \times 10^{-9} C_1(i)^{-1}
\left(\frac{f_{\rm cap}}{100}\right)^{-1}
\left(\frac{d}{100\,{\rm km}}\right)
\left(\frac{q}{5.4 \times 10^7\,{\rm km} }\right)^{-1}
\left(\frac{\rho_{\rm d}}{1 \,{\rm g\, cm}^{-3}}\right)
\,{\rm g\,cm}^{-3},
\end{equation}
where the interior density of bodies, $\rho_{\rm d}$, is assumed to be $1\,{\rm
g\,cm}^{-3}$, the Hill radius of Jupiter is applied to $q$, and
$f_{\rm cap}$ is possibly approximately $100$
\citep{suetsugu11,suetsugu13}. As shown in Figure~\ref{fig4}, $C_1(i)$
is mainly 0.1 to 10. This density is comparable to or less than the
`minimum mass subnebula' disk that contains a mass in solids equal to the
mass of current Jovian satellites and gas according to the solar composition
\citep{canup02}. \rev{It should be noted that the temporally captured
bodies are significantly affected by the central star.  However, the
temporally captured bodies rotate around the host planet, which means that the
perturbation by the central star is roughly canceled out in a temporally
captured orbit. Therefore, the energy loss due to gas drag estimated
above may lead to bound orbits. }

Inclination decreases during the full capture by gas drag, which is
estimated as $C_2(i) = [(\langle \dot i \rangle /i) (e / \langle \dot e
\rangle)]_{e=1}$ in Figure~\ref{fig4}. The
initial inclination is damped during capture for $20^\circ < i <
30^\circ$, while inclinations remain high after capture for other $i$.

However, inclinations keep decreasing due to gas drag after capture.
A dissipation time of the disk, $T_{\rm disk}$,
that is shorter than the damping time of
inclination is thus necessary for the formation of high-inclination
satellites:
\begin{eqnarray}
T_{\rm disk} &\la& i/|\langle \dot i \rangle | \sim f_{\rm cap} T_{\rm K}(q) /
C_2(i)
\nonumber
\\
 &\sim& 7\times 10^3 \left(\frac{f_{\rm cap}}{100}\right)
  \left(\frac{C_2(i)}{0.1}\right)^{-1} \left(\frac{q}{5.4 \times
   10^7\,{\rm km} }\right)^{3/2}
  \left(\frac{M_{\rm p}}{2\times 10^{30} {\rm g}}\right)^{-1/2}\,{\rm yr},\label{eq:tdisk}
\end{eqnarray}
where $M_{\rm p}$ is the host planet mass.
Since the dissipation
processes of circumplanetary disks are not clear yet \citep{fujii}, it is difficult to
discuss the dissipation timescale. However, the dissipation timescale
needed to form
high-inclination satellites seems too short. Therefore, the capture of
high-inclination satellites might have
occurred in the timescale estimated in Equation~(\ref{eq:tdisk})
before the disk dissipation and the resulting satellites tend to have
retrograde orbits (see Figure~\ref{fig4}).

\subsection*{Summary}

I have investigated the time derivatives of orbital semimajor axis $a$,
eccentricity $e$, and inclination $i$ of a body orbiting in a gaseous
disk.

\begin{itemize}
 \item I have derived $\dot a$, $\dot e$, and $\dot i$ for $e > 0.9$ and
       $i < H/2a$ (Equations~\ref{eq:da_he} to \ref{eq:di_he}) and for
       $i > 2H/a$ (Equations~\ref{eq:da_high} to \ref{eq:di_high}). In
       addition, I have modified the formulae derived by AHN;
       Equations~(\ref{eq:mod_adachi_a}) to (\ref{eq:mod_adachi_i}) are valid for $e
       < 0.2$ and $i < H/2a$, where $H$ is the disk scale height.
 \item I have combined the formulae in the limited cases and have
       constructed approximate formulae for $\dot a$, $\dot e$, and
       $\dot i$ (Equations~\ref{eq:da_high} to \ref{eq:di_mid}), which are
       applicable unless $e \geq 1$ or $i > \pi - H/a$.
 \item Using these formulae, I have discussed the orbital evolution of
       satellites captured by a circumplanetary disk. High-inclination
       satellites are formed if the bodies are captured in approximately $10^4$
       years before the disk dissipation.
\end{itemize}

\section*{Competing interests}
The author declares that he has no competing interests.

\section*{Acknowledgements}

I acknowledge the useful discussion with K. Nakazawa, S. Ida, H. Emori,
and H. Tanaka to derive the analytic solutions.
I thank the reviewers for their comments that improved this manuscript.
I gratefully acknowledge support from Grant-in-Aid for Scientific Research (B) (26287101).

\end{document}